\documentclass[aps,prl,amsmath,amssymb,reprint,superscriptaddress,twocolumn,showpacs]{revtex4-1}

 \usepackage{graphicx}
 \usepackage{amsmath}
 \usepackage{amsfonts}
 \usepackage{amssymb}
 \usepackage{pstricks}
 \usepackage{psfrag}
 \usepackage{epsfig}
 \usepackage{changes}
 \usepackage[ansinew]{inputenc}
\usepackage{mathptmx}
\usepackage{helvet}
\usepackage{textcomp}
\usepackage{ulem}

\begin{document}


\title{Hydrodynamic assembly of active colloids: chiral spinners and dynamic crystals}



\author{Zaiyi Shen}
\affiliation{Univ. Bordeaux, CNRS, LOMA (UMR 5798), F-33405 Talence, France}
\author{Alois W\"urger}
\affiliation{Univ. Bordeaux, CNRS, LOMA (UMR 5798), F-33405 Talence, France}
\author{Juho S. Lintuvuori}
\email[]{juho.lintuvuori@u-bordeaux.fr}
\affiliation{Univ. Bordeaux, CNRS, LOMA (UMR 5798), F-33405 Talence, France}

\date{\today}

\begin{abstract}
Active colloids self-organise to a variety of collective states, ranging from highly motile ``molecules'' to complex 3D structures. 
  Using large-scale simulations, we show that hydrodynamic interactions, together with a gravity-like aligning field, lead to tunable self-assembly of active colloidal spheres near a surface.
  The observed structures depend   on the hydrodynamic characteristics: particles driven at the front, {\it pullers}, form small chiral spinners consisting of two or three {\color{black}particles}, whereas those driven at the rear, {\it pushers}, assemble to large dynamic aggregates.  The rotational motion of the puller spinners, arises from spontaneous breaking of the internal chirality. Our results show that the fluid flow  mediates chiral transfer between neighboring spinners. Finally we show that the chirality of the individual spinners controls the topology of the self-assembly in solution: homochiral samples assemble into a hexagonally symmetric 2D crystal lattice while racemic mixtures show reduced hexatic order with diffusion-like dynamics.
\end{abstract}

\maketitle

\section*{Introduction}

Chirality is ubiquitous in Nature, ranging from spiral galaxies to the molecular level. Left- and right-handed species may occur as racemic mixtures, or one of them may be dominant, as for the example the left-twisted trumpet honeysuckle or the dextral helix of DNA. The origin of biological homochirality is not well understood \cite{Bada_Nature_1995,Goldenfeld_PRL_2015}.


The spontaneous emergence of chiral structures~\cite{zhang2016directed,zhang2015toward,kokot2015emergence,Ma_PNAS_2015,gao2013organized,petroff2015fast} are among the most striking aspects of the collective behavior of motile bacteria and synthetic microswimmers~\cite{mike2012review,moran2017phoretic,whitesides2002self,chantal2017review}.  
Synthetic microswimmers can be realised by active Janus particles~\cite{moran2017phoretic}, where the self-propulsion results for example from surface forces generated by catalytic activity, light absorption, or induced-charge electro-osmosis (ICEO). In many cases the particles  settle on a solid surface and their axes weakly align on a gravitational or magnetic field. The observed self-assembled states range from dynamic clusters \cite{palacci2013living,ginot_2018_aggregation-fragmentation,zhang2016directed,linek2012rotors} and chains~\cite{yan2016reconfiguring} to spirals~\cite{zhang2016natural} and chiral colloidal molecules which rotate rapidly around their axis~\cite{zhang2016directed,zhang2015toward,Ma_PNAS_2015,gao2013organized}.

The interparticle interactions which are at the origin of self-organised structures, are not well understood. Noteworthy exceptions are provided by catalytic Janus spheres with a hydrophobic hemisphere~\cite{gao2013organized}
, and particles driven by AC electric fields which result in strong multipole interactions~\cite{Ma_PNAS_2015,zhang2016directed,zhang2015toward}.
Little is known, on the mutual forces of particles which propel through thermal or chemical surface forces: Besides the interactions mediated by their gradient fields, hydrodynamic forces are always present and have been shown to influence particle aggregation and dynamics~\cite{marchetti2012review,evans2011orientational,bricarde2013emergence,zottl2016emergent,schaar2015detention,alarcon_2013_spontaneous,delmotte_2015_large-scale,alarcon_2017_morphology,ishikawa_2008_coherent}.  Specifically, hydrodynamic flow has been shown to lead the formation of complex structures such as lines~\cite{thutupalli_2018_flow-induced}  and crystals~\cite{thutupalli_2018_flow-induced,singh2016universal}, when the self-propelled particles are confined by surfaces. Both the gradient and the flow fields depend on the activity difference between front and back hemispheres of the Janus colloid; unfortunately, the details of the interactions remain unclear and in most experimental cases not even the squirmer characteristics (puller or pusher) are known.


\begin{figure*}[t]
\centering
\includegraphics[width=2.0\columnwidth]{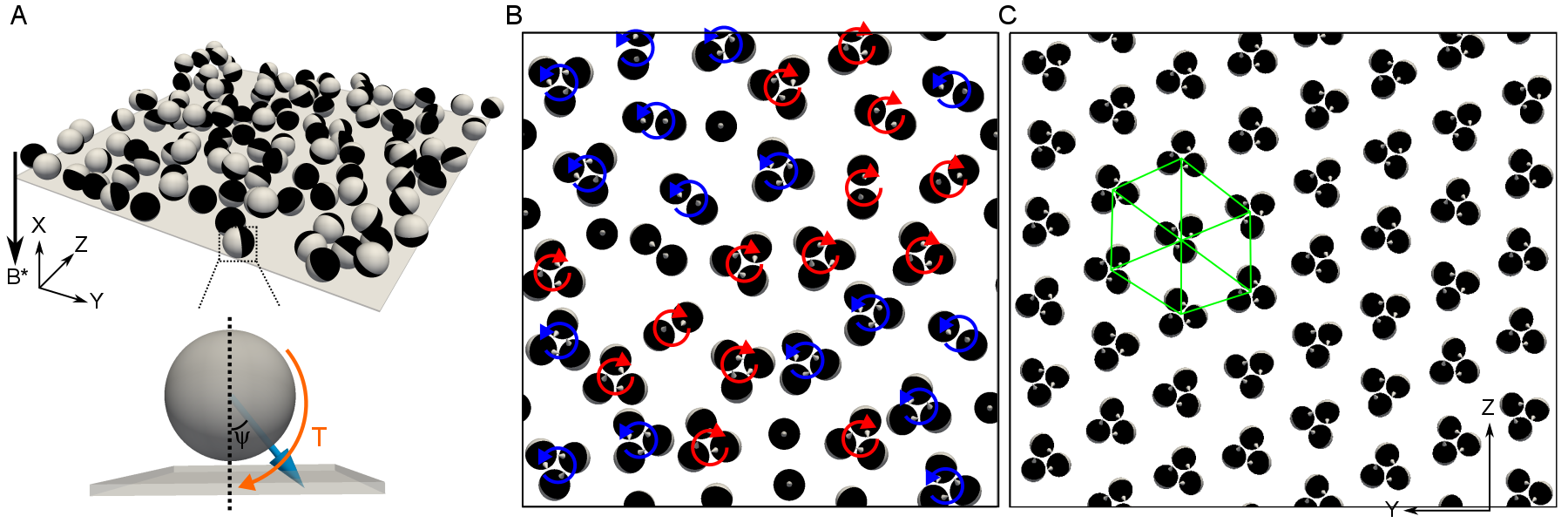}
\caption{{\bf Hydrodynamic self-assembly of active colloids near a surface.} ({\bf A}) Schematic of the system: The blue arrow indicates the particle axis and the black arrow shows the direction of the external field. The external field $B^*$ creates a torque (orange arrow) which rotates the particles towards the solid surface. Examples of the observed self-assembled states: ({\bf B}) Starting from a gas phase, strong pullers assemble into hydrodynamically bound dimers and trimers, which spontaneously break internal chiral symmetry, leading to a spinning motion. ({\bf C}) A homochiral solution forms  a stable 2 dimensional crystal, where the spinners are arrested into a hexagonal lattice. (The simulations in ({\bf B}) we carried out using $N = 90$ particles, corresponding to an area fraction $\phi\approx 31$\% with $\beta = +5$ and $B^* = 2.0$ (see text for details), and in ({\bf C}) $N = 144$ ($\phi\approx 26\%$) with $\beta = +5$ and $B^* = 1.5$). 
 \label{model}}
\end{figure*}

This paper aims at elucidating the 
role and importance of hydrodynamic interactions in the formation of dynamic structures
, and in particular the spontaneous emergence of chiral aggregates. Using hydrodynamic simulations, we study a solution of active colloids, modelled as spherical squirmers with a radius $R$, confined between two flat walls (see methods). The particles interact due to hydrodynamics and  are subject to a gravity-like field turning their axis towards a confining wall (Fig.~\ref{model}A). As a most striking feature we find that achiral particles spontaneously aggregate into chiral clusters which spin around their axis (Fig.~\ref{model}B and movie 1) and can transfer their chirality between nearby spinners via hydrodynamic field. The overall hydrodynamic interactions between the spinners are repulsive, leading to the formation of a 2D crystal of homochiral spinners (Fig.~\ref{model}C), where the spinners are arrested in a hexagonally symmetric lattice, while racemic mixture shows diffusion-like behaviour with reduced hexatic order. The observed structures can be further tuned using the details of the self-generated flow field: chiral clusters are formed only by pullers due to directional hydrodynamic interactions~\cite{supplement}, while pushers form large ``ferromagnetic'' aggregates due to mutual hydrodynamic interactions~\cite{supplement}.  
\section*{Results}

\begin{figure*}
\centering
\includegraphics[width=2\columnwidth]{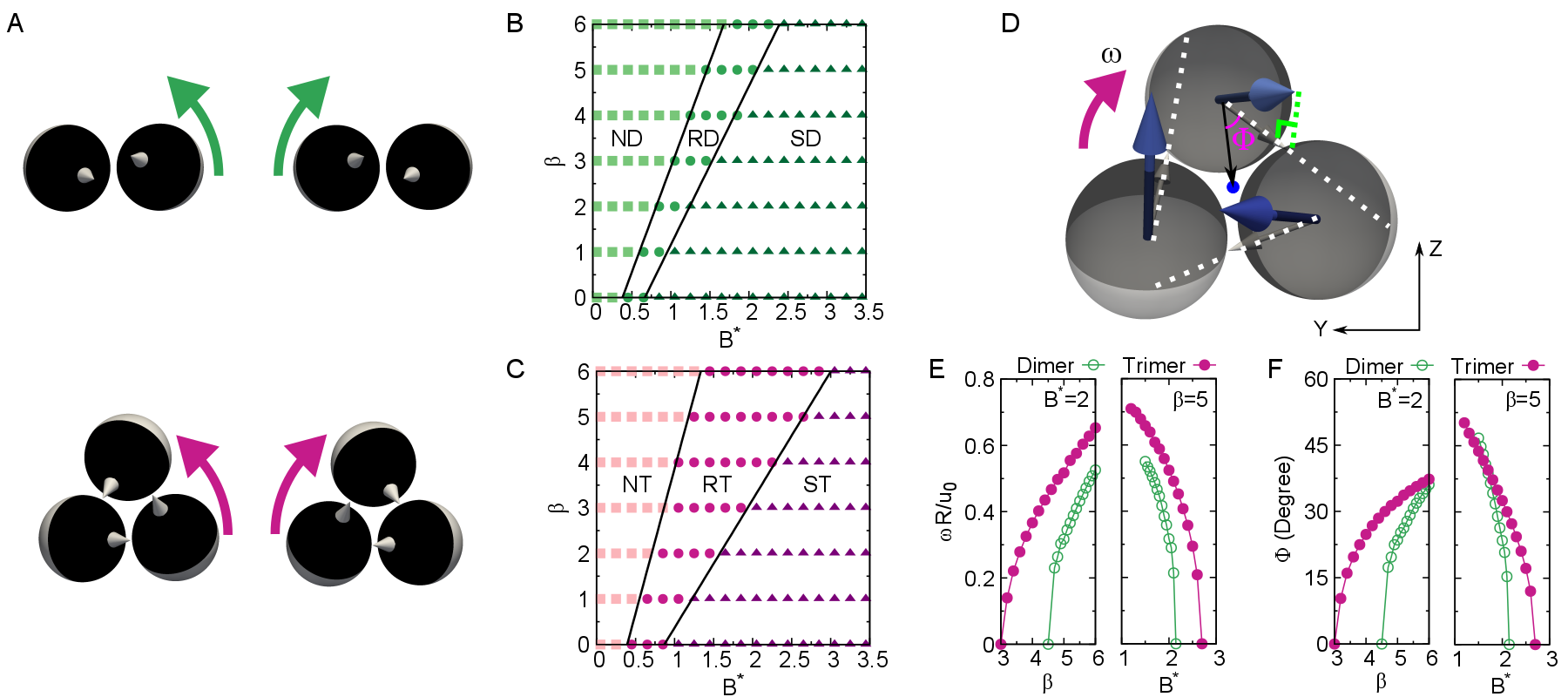}
\caption{{\bf Rotational motion of the puller dimers and trimers.} ({\bf A}) The internal chiral structures of the dimer and trimer spinners (bottom view). The green (violet) arrows show the spinning direction of the dimers (trimers): clockwise twisting clusters spin counterclockwise and vice versa. ({\bf B}) Phase diagram for dimer spinners by varying $\beta$ and $B^*$.  Rotating Dimers (RD) appear between the two solid lines, while stationary dimers occur in the region (SD) and bound states do not exist in the region (ND). ({\bf C}) The same for trimers consisting of three particles. ({\bf D}) Schematic 3D view of a clockwise rotating trimer. The blue arrows show the particle axes, and the white dashed lines present their projection in the wall plane. The angle $\Phi$ defines a measure of internal chirality.  ({\bf E}) Angular velocity of the spinners $\omega$ as a function of $\beta$ ($B^*=2$) and $B^*$ ($\beta=+5$). ({\bf F}) Chirality $\Phi$ as a function of $\beta$ and $B^*$.}
\label{rotor}
\end{figure*}

\subsection*{Achiral active particles form small chiral spinners}

From theoretical calculations~\cite{ishimoto13} and lattice Boltzmann simulations~\cite{shen18}, it is known that isolated pullers can be hydrodynamically trapped by a solid boundary. We characterise the dynamics and stability of small puller clusters, consisting of two or three particles, near a solid surface in the presence of an aligning field $B^*$ normalised by the magnitude of the hydrodynamic (viscous) torque arising from the self-propulsion (see methods for details) .
The dynamical states of two or three pullers depend on the squirmer parameter $\beta$ and $B^*$ (Fig.~\ref{rotor}).
For weak fields there is no bound state (region ND and NT in the Fig.~\ref{rotor}B and C); when initialised as dimer or trimer, the particles rapidly separate. At slightly larger fields, we observe the rapid formation of chiral clusters (region RD and RT in Fig.~\ref{rotor}B and C) which rotate clockwise or counterclockwise (Fig.~\ref{rotor}A).  These rotating dimers and trimers are bound through their mutual flow fields (Supplementary Fig.~S3); the inclination $\psi$ of the particle respect to wall normal (Fig.~\ref{model}A) is finite and does not reach the minimum-energy state $\psi=0$ favored by the applied torque.

The internal structure of the spinning dimers and trimers spontaneously breaks chiral symmetry (Fig.~\ref{rotor}A), and the handedness of the cluster determines the spinning direction: clockwise twisted aggregates spin counterclockwise and vice versa. 
To measure the amount of internal chirality, we define an angle $\Phi$ between the line from the particle to the geometrical center of the spinner and the projection of the particle {\color{black} axes} on to the plane parallel to the surface (Fig.~\ref{rotor}D).
The chirality $\Phi$ and the angular  velocity $\omega$ are connected: both increase when $\beta$ is increased and decrease upon increasing $B^*$ (Fig.~\ref{rotor}E and F).

The internal chirality of the cluster is the requirement for the overall spinning motion. 
The rotation stops when $\Phi=0$ (Fig.~\ref{rotor}E and F). Beyond the threshold value we observe dimers and trimers of zero angular velocity when starting from appropriate initial conditions (SD and ST in Fig.~\ref{rotor}B and C, respectively). Generally, the hydrodynamically bound puller states require a slight inclination of the particle axis ($\psi>0$); very strong fields $B^* > 5.0$ impose $\psi=0$ and the interactions become repulsive (Supplementary Fig.~S3).

\begin{figure*}
\centering
\includegraphics[width=1\textwidth]{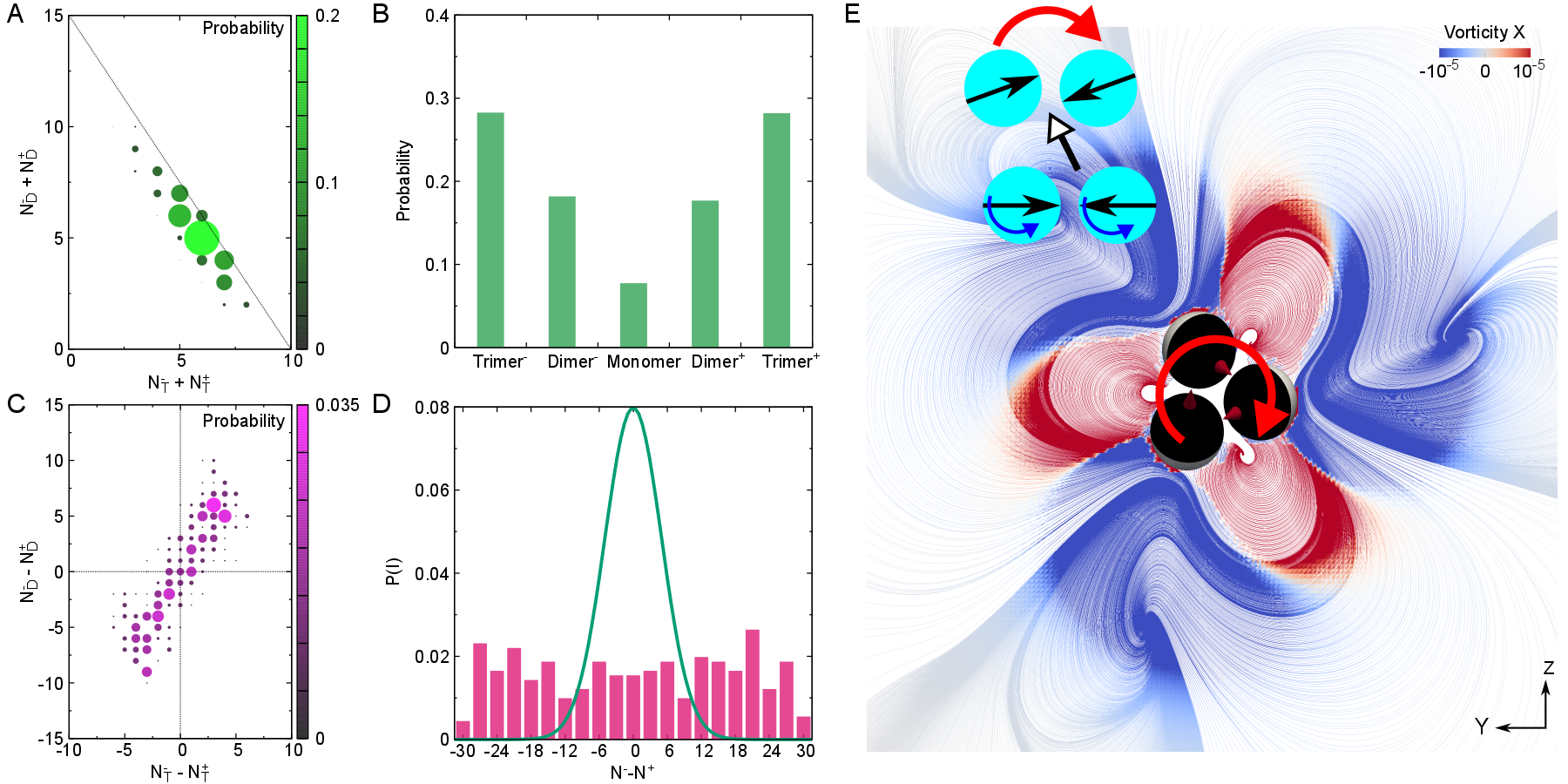}
\caption{{\bf Chiral transfer between the spinners.} ({\bf A}) The total number of dimers $N_D^-+N_D^+$ as a function of the total number of trimers $N_T^-+N_T^+$ . The dashed line corresponds to states without monomers. ({\bf B}) The probability that a given particle joins to a left-handed or right-handed trimer or dimer, or remains on its own.  ({\bf C}) The number difference of counterclockwise and clockwise spinning dimers, $N_D^- - N_D^+$, as a function of that for trimers $N_T^- - N_T^+$. Their strong correlation indicates rotational coupling between dimers and trimers. ({\bf D}) The probability distribution function $P(I)$ of the number difference of particles in counterclockwise spinners and clockwise spinners, $I = N^--N^+$ with $N^\pm=2N_D^\pm+3N_T^\pm$. The solid line shows a Gaussian distribution as expected for uncorrelated spinners~\cite{supplement}. ({\bf E}) Observed flow field  $\mathbf{v}(\mathbf{r})$ produced by a  clockwise rotating trimer in the plane across the center.
The color map indicates the vorticity $\nabla\times \mathbf{v}$.  Beyond the immediate vicinity, the vorticity is negative (blue), thus it rotates the particles of a dimer counterclockwise (turquoise spheres in the inset), leading to a chiral symmetry breaking within the dimer, which aligns its spin on that of the trimer. (The simulations in ({\bf A} - {\bf D}) we carried out with $N=30$ particles ($\phi \approx 30$\%) using $\beta = + 5$ and $B^*=2.0$). }
  \label{chiral}
  \end{figure*}

\subsection*{Chiral transfer via hydrodynamic coupling}

To study the {\color{black} collective dynamics} of the spinner phase, we implemented 303 statistically independent simulations of 30 particles each (giving an area fraction of approximately 30\%), with parameters {\color{black}$\beta=+5$ and $B^*=2.0$}.  All systems were initialised randomly in space but all the particles pointing directly at the wall~\cite{supplement}, to ensure that the initial configurations had no chirality. Most particles rapidly join into clusters to form either a dimer or a trimer (Fig.~\ref{chiral}A). The overall statistics of the particles forming a clockwise or counterclockwise turning spinners, show a symmetrical distribution (Fig.~\ref{chiral}B) when averaged over all simulations, signaling on average a racemic mixture.

To analyze the global chirality we calculate the imbalance $I=N^--N^+$ between the counterclockwise and clockwise spinners.  In most cases the imbalance rapidly reaches a stationary value (Supplementary Fig. S6).
For non-interacting spinners, one expects a Gaussian probability distribution $P_G(I)= e^{-I^2/2\Delta I^{2}}/ \sqrt{2\pi\Delta I^2}$. Yet from the simulations we obtain an approximately flat distribution (Fig.~\ref{chiral}D), implying a `ferromagnetic' interaction between the `spins' of nearby clusters. In the simplest mean-field picture with a positive coupling constant $\lambda$, such a coupling modifies the distribution according to $P(I)\propto e^{-{I^2}/{2\Delta I^2} + \lambda I^2}$, {\color{black} which, with $\lambda\sim 1/\Delta I^2$, would account for the distribution shown in Fig.~\ref{chiral}D.}

Our simulation data further proves the existence of a spin-dependent interaction between dimers and trimers. We observe a strong correlation between their imbalance parameters, $I_D=N_D^--N_D^+$ and  $I_T=N_T^--N_T^+$ (Fig.~\ref{chiral}C), with a coefficient $\rho(I_D,I_T)=cov(I_D,I_T)/\sqrt{var(I_D)var(I_T)} \approx 0.8$, revealing a transfer of chirality between trimers and dimers.

This chiral coupling results from hydrodynamic interactions, and its ferromagnetic nature can be understood by considering the flow field of spinning trimer, more precisely its vorticity (Fig.~\ref{chiral}E). At moderate or large distances from the center, the flow of a right-handed trimer has negative vorticity, inducing left-handed rotation in the fluid (blue area in the Fig.~\ref{chiral}E). As a consequence, this flow rotates the particles of a dimer counterclockwise, thus breaking its chiral symmetry and leading to a clockwise spinning dimer.

\subsection*{Hydrodynamic interactions and crystallization}

\begin{figure*}[tbh!]
\centering
\includegraphics[width=2.0\columnwidth]{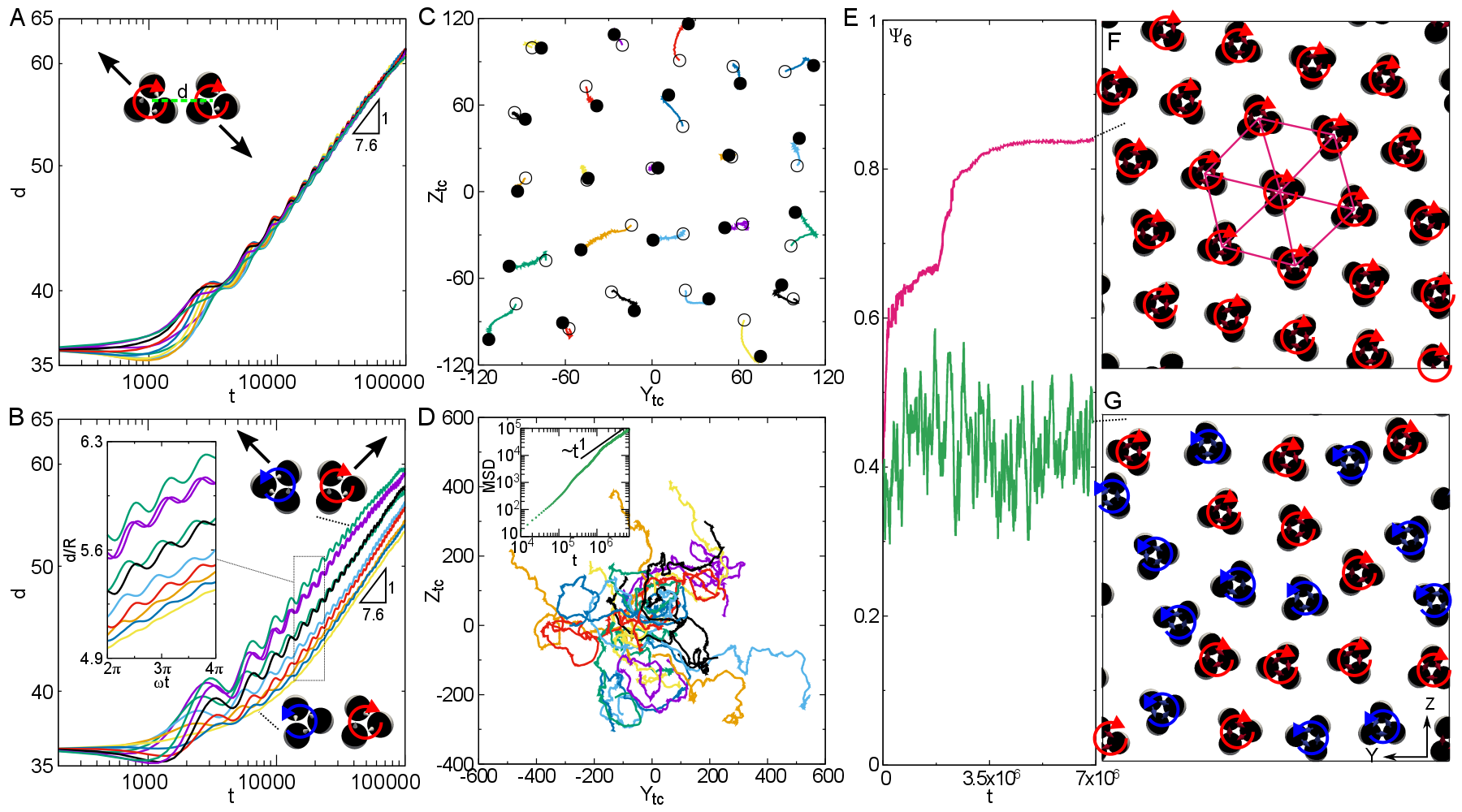}
\caption{{\bf Hydrodynamic interactions and self-assembly of spinners rotating the same (top row) and opposite (bottom row) directions.} ({\bf A},{\bf B}) The distance $d$ as a function of time $t$ for a pair of trimers rotating in the ({\bf A}) same and ({\bf B}) opposite directions. The lines correspond to different initial orientations of the spinners (see text for details). The data is fitted with a power law $\log d\propto\kappa \log t$, with $\kappa \approx \frac{1}{7.6}$. The black arrows give the direction of translation of the spinners and the inset in ({\bf B}) shows the $d(t)$ over one period of rotation. ({\bf C},{\bf D}) The trajectories of the centre of mass coordinates for 24 spinners: ({\bf C}) Homochiral spinners quickly organise into a crystal with a hexagonal symmetry (the open (closed) symbols show the initial (final) positions), while ({\bf D}) racemic mixture shows diffusion-like behaviour, with a mean square displacement (MSD) scaling linearly with time $t$ at long times (inset). ({\bf E}) The time evolution of the hexagonal order parameter $\psi_6=\langle e^{i6\psi_{ij}}\rangle$ for the monochiral spinners (red curve) and racemic mixture (green curve). The snapshots from the the steady state: ({\bf F}) monochiral spinner phase showing a stable hexagonal order and ({\bf G}) frustrated order of the racemic mixture. (The simulations in ({\bf C}-{\bf E}) we carried out with $N=72$ particles in a box with square cross section $L=30R$, corresponding to an area fraction $\phi \approx 25$\%, using $\beta = +5$ and $B^* = 2.0$.)}
\label{HI}
\end{figure*}

To study the formation of super structures, we characterise the hydrodynamic interactions between the spinners (Fig.~\ref{HI}). We use pullers with $\beta= +5$ and $B^*=2.0$, where the triplet state is stable (see Fig.~\ref{rotor}). 

The slip velocity of an active Janus particle as a function of the polar angle $\alpha$ reads as $u(\alpha)=\tfrac{3}{2}u_0 \sin\alpha (1 + \beta \cos\alpha)$; in the absence of confinement, the corresponding far-fields decay as $r^{-3}$ and $\beta r^{-2}$. A nearby wall modifies the velocity field. An active particle with $\beta=0$ results in an {\it inward} radial velocity $\propto r^{-4}$ along the solid boundary, as obtained from the method of reflexion~\cite{Mor10}.
Our simulations show that a puller ($\beta>0$) creates an {\it outward} flow (Supplementary Fig.~S3).
The superposition of the Janus colloids of a trimer spinner, results in an outward flow along the boundary plus a modulation of threefold symmetry, $v=C r^{-n} + C_3 \cos(3\theta)r^{-m}$.

We measure the time evolution of the centre-to-centre distance $d$ of a pair of spinners turning the same (Fig.~\ref{HI}A)  and opposite (Fig.~\ref{HI}B) directions. All curves start at an initial separation $d_0\approx 4.5R$ but with different initial orientations (inset in Fig.~\ref{HI}B). We observe two main features: an overall increase of the distance $d$ with time, and a periodic modulation (Fig.~\ref{HI}A and B).

The overall increase of $d$ indicates hydrodynamic repulsion, or mutual outward advection of the trimers, in qualitative agreement with the flow field discussed above. We identify the (angle-averaged) velocity field $C/d^n$ of one trimer with the advection velocity $\dot d$ of its neighbor, where the dot indicates the time derivative. The resulting differential equation $\dot d = C/d^n$ is readily solved, $d\propto t^{\frac{1}{n+1}}$. From the fits in Fig. \ref{HI} A and B, we find $n\approx 6.6$.  This non-integer power law probably results from the finite box size of the present simulations, the vertical height of which corresponds to six particle radius (see methods).

Both Figs. \ref{HI} A and B show a modulation with angular frequency $3\omega$, which is three times the spinning frequency $\omega$ of a trimer. The inset in Fig.~\ref{HI}B displays the occurrence of three maxima of $d(t)$ during one period of rotation of the spinners. This can be understood in terms of the angular variation of the flow field of an individual trimer (Fig.~\ref{chiral}E), which shows a strong contribution of threefold symmetry, $C_3 \cos(3\theta)/d^m$. Thus in addition to the mean repulsion, each trimer experiences a periodic back-and-forth advection in the velocity field of its neighbor.

There are two main differences between pairs spinning in the same or opposite directions: The latter (Fig. \ref{HI}B) shows a larger modulation amplitude, and in the initial state the repulsive force depends strongly on the relative orientation of the pair: The lower curve in Fig. \ref{HI}B is calculated for trimers with the same initial orientation ($\Delta\theta_{12}^0=\theta_1^0-\theta_2^0=0$), whereas the upper curve describes an out-of-phase pair ($\Delta\theta_{12}^0=\pi/3$); the intermediate curves interpolate between these values. This can be understood by expanding the 3-fold flow field of one trimer in terms of the orientation angle of the neighbor, resulting in $v_{\mathrm{phase}} \propto C_p \cos(3\Delta\theta^0_{12})$. Its opposite sign for in-phase and out-of-phase pairs results in different advection velocities, as observed in Fig. 4B. The corresponding angular velocity favors a cogwheel type motion and phase-locking of nearby spinners.

A striking difference occurs in the phase behavior of homochiral and racemic systems. Fig.~\ref{HI}C and D show the evolution of the center of mass coordinates of 24 spinners in a box with a square surface area: Starting from random initial positions on a plane near the surface, the homochiral system quickly organise into a 2D hexagonal crystal (Fig.~\ref{HI}C), with a steady-state order parameter $\psi_6=\langle e^{i 6 \psi_{ij}}\rangle \approx 0.82$ (red curve in Fig.~\ref{HI}E) and where all spinners rotate at nearly identical frequencies.  Similar results have been obtained for systems of variable particle number and rectangular surface area (see {\it e.g.} Fig.~\ref{model}C).

In the case of the racemic mixture, the spinners do not settle in a crystalline state. Their centre of mass trajectories show diffusion-like behavior at long times (Fig.~\ref{HI}D) and the order parameter $\psi_6$ is lower and does not reach a steady value (green line in Fig.\ref{HI}E). The system retains a reasonably high value of hexatic order but shows a frustrated state  (Fig.~\ref{HI}G). From the above discussion of the hydrodynamic pair interactions, we are led to the conclusion that the frustrated order and the diffusion-like behavior arise from the velocity component $v_{\mathrm{phase}}$  which favors cogwheel-like motion of nearby spinners of opposite chirality.  We also observe decoupling of the time-scale of the rotation of an individual spinner and the diffusional time scale $\tau_{D} = R^{2}/D \approx 2\tau_{spinner}$ (see also movie 2 initially played with $0.1\tau_{spinner}$ between the frames and 10 frames per second and then accelerated $40$ times.)


\subsection*{Tunable motile pusher aggregates}

\begin{figure*}[t!]
\centering
\includegraphics[width=1\textwidth]{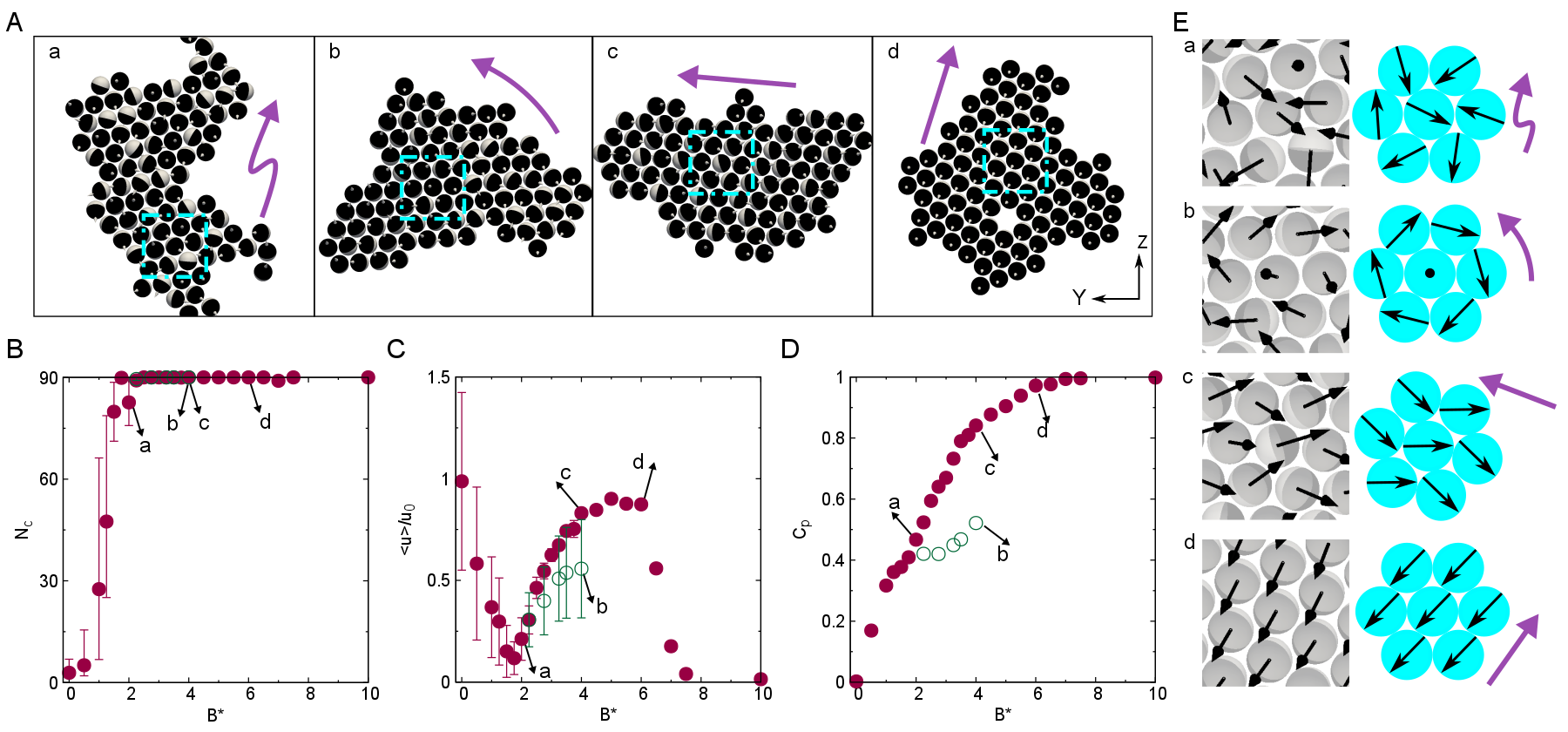}
\caption{{\bf Cluster dynamics of pushers tuned by the external field $B^*$}. ({\bf A}) Upon increasing $B^*$ we observe a sequence of dynamic states: random motion (a), rotation (b), linear translation (c and d). ({\bf B}) Mean cluster size $N_c$ as a function of $B^*$.  The error bar gives the maximum and minimum size of the cluster. ({\bf C}) Average velocity $\langle u\rangle/u_0$ normalised by the free swimming speed $u_0$ of a single particle. Error bars give the standard deviation. ({\bf D}) The alignment coefficient $C_p=\left< \cos\alpha_{ij}\right>$ of the particles. ({\bf E}) Observed swimmer orientations in the dynamic states (a-d) in A. The simulation images are from dashed frames in ({\bf A}). (The simulations were carried out using $N = 90$ ($\phi\approx 30$\%) with $beta = -5$.)}
\label{pusher}
\end{figure*}

Finally we turn to the dynamics of pusher ($\beta < 0$) clusters. 
For moderate values of the squirming parameter $\beta \sim -1$ and aligning field $ B^*\sim 1$, we observe a mixture of rotating and translating clusters with random internal orientation, which can break and reform (right panel in movie 3), and are reminiscent of what is observed with phoretic swimmers near a no-slip surface~\cite{ginot_2018_aggregation-fragmentation}.  For large $B^*$ the hydrodynamic interactions are attractive and favour clustering~\cite{singh2016universal} (See supplementary Fig.~S3). In the case of strong pushers, all the particles aggregate into a single cluster with a hexagonal packing of the particles (see Fig.~\ref{pusher}A for $\beta= -5$ and $B^* > 2$).

Depending on the strength of $B^*$, the mutual alignment of the strong $\beta = -5$ pushers produces motile structures which exhibit erratic motion, rotation or translation (Fig.~\ref{pusher}A and movies 4, 5 and 6). Moreover,  the alignment leads to an increase of the cluster velocity (Fig.~\ref{pusher}C) and is easily observable from the increase of the order parameter $C_p=\left< \cos\alpha_{ij}\right>$, where $\alpha_{ij}$ is the relative angle between the axes of squirmers $i$ and $j$, and where angular brackets denote the average over all pairs (Fig.~\ref{pusher}D).
For a low aligning field, a slowly and randomly moving cluster is formed (state a in Fig.~\ref{pusher}A and C).

This dynamic results from the changing and erratic arrangement of the swimmer orientations (state a in Fig.~\ref{pusher}E).
A bifurcation is observed in the range $2<B^*<4$ (see Fig.~\ref{pusher}C and Fig.~\ref{pusher}D), where two metastable states b and c show circular and zigzag arrangements of the particle orientations (state b and c in Fig.~\ref{pusher}E), resulting in rotating and translating clusters, respectively. Increasing $B^*$ further, a completely aligned state d is observed, moving at a velocity close to that of a single swimmer (see Supplementary movie 7). Very high fields ($B^*>7$) orient the particles perpendicular to the surface, and the clusters become immobile (Fig.~\ref{pusher}C).

\subsection*{Phase diagram of collective squirmer dynamics near a surface}

\begin{figure*}[!th]
\centering
\includegraphics[width=2.0\columnwidth]{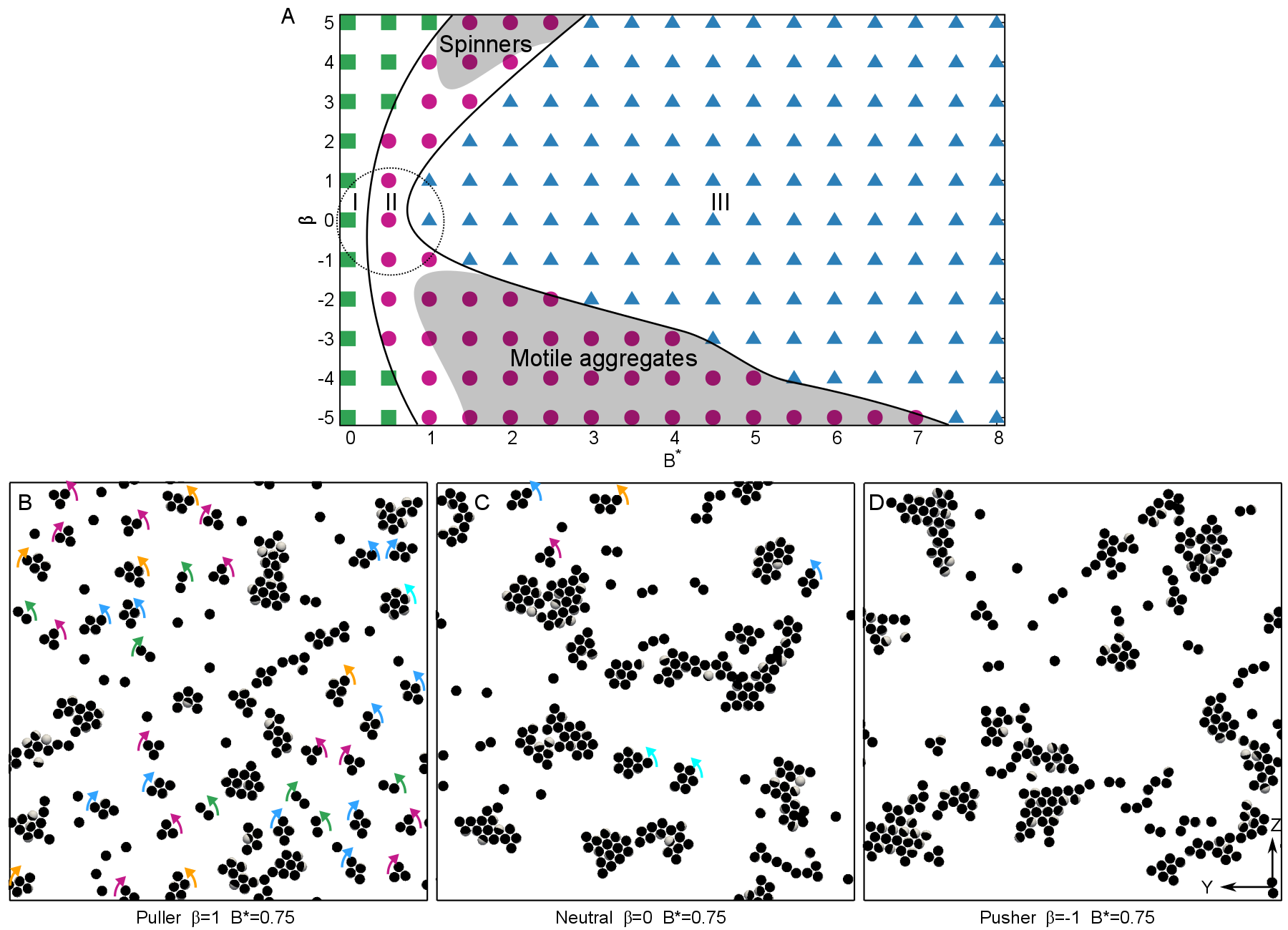}
\caption{{\bf (A) Detailed phase diagram in the $(B^*,~\beta)$-space:} gas-like state (green squares), dynamic cluster state (red circles) and stationary clusters or isolated particles (blue triangles). The two shaded regions mark the ranges where we observe only rotating dimers and trimers for $\beta > 0$ or a single motile aggregate for $\beta < 0$. Between these two ranges, mixture of clusters of different sizes is observed. Examples of observed states with low squirmer parameter $|\beta| \leq 1$ and aligning field $B^*=0.75$: ({\bf B}) Weak $\beta = +1$ pullers form hydrodynamically bound small chiral spinners and larger mobile aggregates. ({\bf C}) Neutral $\beta = 0$ squirmers form larger motile aggregates and small number of rotating chiral clusters, while ({\bf D}) weak $\beta = -1$ pushers assemble into motile aggregates which can break and reform. (In ({\bf A}) the simulations were carried out with 90 particles, corresponding to an area fraction $\phi\approx 31$\% and in ({\bf B-D}) with $N=300$ corresponding to $\phi\approx 15\%$.)
\label{phase}}
\end{figure*}

To generalise our results, we have calculated a detailed phase diagram of $N=90$ particles in the $\beta-B^*$ space (Fig.~\ref{phase}A).
For a very weak field, the swimmers are not bound to the lower wall, but move randomly between the two confining surfaces, leading to a gas-like state (I) (squares in Fig.~\ref{phase}A and supplementary Fig.~S5A-C). When increasing the strength of the external field, the particles swim near the surface, and spontaneously form a variety of dynamic structures (Supplementary Fig.~S5D-F). The system can be tuned such that a pure spinner phase consisting primarily dimers and trimers or a single aggregate is observed, for pullers and pushers, respectively (shaded regions in Fig.~\ref{phase}A).
Finally, for a strong external {\color{black} field}, stationary clusters are observed (triangles in Fig.~\ref{phase}A). The size and shape of these clusters depend on the squirmer parameter $\beta$ and the strength of the field (Supplementary Fig.~S5G-I).

In typical experimental realisations of active Janus particles~\cite{moran2017phoretic}, the squirmer parameter is thought to be reasonably low $|\beta| \leq 1$. For weak $\beta = + 1$ pullers, chiral spinners are quickly formed and they co-exist with larger aggregates (Fig.~\ref{phase}B and movie 3). Decreasing $\beta$ towards pushers, the spinners become unstable and in the case of neutral $\beta=0$ squirmers only few rotating clusters remain (Fig.~\ref{phase}C and movie 3) and most of the particles are hydrodynamically bound to form motile clusters consisting of $\gtrsim 10$ particles. Weak $\beta = -1$ pushers form motile aggregates which can fragment and reform during the simulations (Fig.~\ref{phase}D and movie 3), similarly to what is observed with phoretic swimmers near a no-slip surface~\cite{ginot_2018_aggregation-fragmentation}.

\section*{Discussion}

We provide a generic route for a tunable structuration of microswimmers, based on hydrodynamic interactions in the presence of simple guidance by an aligning field. When deposited on a surface, the interactions between the particles, arising from the self-induced flow field, leads to formation of chiral spinners and tunable dynamic crystals.  The coupled rotation of the self-assembled spinners demonstrates a chiral transfer mediated by hydrodynamic interactions. Finally we show that homochiral spinners can self-organise into stable 2 dimensional crystal lattices, providing a hydrodynamic route to multiscale self-assembly of active particles.

\subsection*{Spontaneous symmetry breaking and chiral spinners}

After starting simulations of two or three pullers at moderate field from an achiral initial state, we rapidly observe formation of a spinning dimer or trimer. Left-handed and right-handed spinners occur with equal probability. Their handedness is related to the twist of the particles' axes with respect to the center (Fig.~\ref{rotor}A), and the spinning direction corresponds to what is expected for particles propelling along their axes (white arrows in Fig.~\ref{rotor}A). There is no bound state for $B^*<0.3$ (Fig. \ref{rotor}B and C), indicating that the underlying hydrodynamic force is attractive only for sufficient strong inclination of the particle axes towards the wall.

Formation of chiral spinners was observed in several experiments on Janus colloids driven by AC electric  or magnetic fields \cite{Ma_PNAS_2015,zhang2016natural,kokot2015emergence}, or chemical reactions \cite{gao2013organized}. The active particles are mostly pushers, whereas in our simulation only pullers form stable spinners. This is not surprising, as the attractive forces are very different. In the experiments~\cite{Ma_PNAS_2015,zhang2016natural,kokot2015emergence}, the self-assembly is due to induced-charge, magnetic-dipole, and dispersion forces, which are independent of the squirmer parameter $\beta$, whereas our simulations rely on hydrodynamic interactions only, which depend on the sign of $\beta$. 

\subsection*{Chirality transfer}
Our simulations show two mechanisms of chiral transfer mediated by hydrodynamic interactions. First, at very short distances, the flow field of one spinner drags the nearby part of its neighbor, which results in cogwheel-like phase locking and rotation of the two spinners in opposite directions. This behavior was observed experimentally pairs of externally driven star-shaped microrotors~\cite{leonardo2012hydrosynchro} and for neighboring hexagonal clusters of phoretic swimmers~\cite{Aub2018}.

The second mechanism of chiral transfer arises from the vorticity of the far-field of one spinner, which slightly rotates any nearby objects in the opposite direction. This is a weak effect and of little consequence for nearby monomers. If there is, however, a freshly formed cluster which has not yet broken chiral symmetry, the vorticity will rotate each of its components and thus favors the formation of a spinner rotating in the same direction, as illustrated in Fig. \ref{chiral}E for a nearby dimer.

Our 303 independent simulations of 30 active particles show, in the steady state, a strong coupling between the spinning directions of dimers and trimers (Fig.~\ref{chiral}C). The initial stage of the spinner formation strongly suggests that flow-mediated chiral transfer is the underlying mechanism. Indeed, it is observed that a first-formed trimer spinner can impose its chirality on an achiral dimer (See supplementary movie 8). 

In a recent experiment~\cite{zhang2016natural},  a similar phenomenon was observed for a pair of rotating three-armed spirals consisting of about 15 active Janus particles. Initially of opposite chirality and showing cogwheel-like motion, one spiral suddenly changes its configuration and chirality, and synchronises its rotation with its neighbor.

\subsection*{Hydrodynamic interactions and spinner crystals}

Our results show that the pair interactions between the spinners is repulsive, leading to non-vanishing hexagonal order near a surface. When the spinners are co-rotating, the hydrodynamic repulsion is isotropic. The situation is markedly different for a pair of spinners rotating opposite directions. Now we observed strong dependence of the initial orientation between the pair to the repulsions. Further, the pair translates along a common axis, perpendicular to the repulsion, as have been theoretically predicted for oppositely spinning active rotors~\cite{leoni_2010_dynamics}.

These differences in the pair-interactions and dynamics, leads to stark differences in the phase behaviour. In the homochiral system a self-assembly into a 2D crystal is observed. The spinners are arrested on a hexagonally symmetric lattice. The racemic mixture sustains reduced hexatic order, and the dynamics becomes diffusion-like at long times. Remarkably this dynamic arises solely from the hydrodynamic mixing resulting from the self-generated flow fields of the individual particles. The importance of the active flow fields is provided when the results are compared to the driven 2D rotors, where the racemic mixture is expected to form a lamellar phase~\cite{yeo_2015_collective}. In our case the monochiral spinners show slightly stronger repulsion compared to the oppositely turning spinners. This could be assumed to favour mixing of the species. Indeed our results with the racemic mixture supports this hypothesis (See supplementary movie 2).

\subsection*{Experimental relevance}

Our simulation results depend strongly on the external field $B^*$ and on the squirmer characteristics $\beta$ of the active particles. We discuss realistic values of these parameters and compare our findings with recent experiments.

Gravity and a vertical magnetic field are possible realisations of the external field $B^*$ turning the squirmers towards the confining wall. Active Janus colloids consisting of a silica or polystyrene sphere with a metal cap on one hemisphere, are subject to a gravitational torque due to the large density of the cap. For a 50 nm gold layer on a sphere of radius $R=1\mu$m, one readily calculates the orientational potential energy  $W=W_0\cos\psi$ with $W_0=0.5\times10^{-19}$~J; the same order of magnitude is achieved for the magnetic energy of ferrite particles. The resulting torque on the particle axis is comparable to the viscous torque at velocities $\sim 1\mu$m/s, thus corresponding to a dimensionless field $B^*\sim 1$, which in view of Fig.~\ref{phase}A is the most interesting range of parameters.

The squirmer parameter $\beta$ is unknown for most chemically fueled microswimmers, and may take either sign for those driven by induced-charge electroosmosis. On the contrary,  laser-heated Janus particles which self-propel due to thermoosmosis-driven are expected to be pullers: The large heat conductivity of metals reduces the slip velocity on the active cap \cite{Bic_PRE_2013} and leads to a positive value of $\beta$ \cite{Wue_PRL_2015}; a sufficiently thick cap forms an isotherm  with $\beta=5$. Thus laser-heated swimmers should be  good candidates for the observation of chiral clusters bound by hydrodynamics.

Our findings for $\beta<0$ compare rather well with the cluster dynamics observed for chemically active Janus colloids. For moderate values of $\beta$ and $ B^*$, we observe a mixture of rotating and translating clusters with random internal orientation, which can break and reform (Fig.~\ref{phase}D and right panel in movie 3), and which are similar to recent experiments with phoretic swimmers near a surface~\cite{ginot_2018_aggregation-fragmentation}.  At higher values of the external field $B^*$ and the squirming parameter $\beta$, we find large ``ferromagnetic'' aggregates, which are reminiscent of the living crystals observed with weakly magnetic swimmers near a surface~\cite{palacci2013living}.

Finally, we address the effects of rotational diffusion, which is not accounted for by our simulations and which would result in fluctuations of the particle axis according to the probability distribution function $P(\psi)\propto e^{-W/k_B T}$. Since the potential energy scale $W_0$ is about ten times larger than the thermal energy, these fluctuations are small and we may safely assume that they would not change the qualitative results on the formation of chiral colloidal molecules nor affect the phase diagram of Fig.~\ref{phase}.

\section*{Materials and Methods}
%

We use lattice Boltzmann method to simulate the active colloids~\cite{supplement}, by employing a classical squirmer model~\cite{lighthill52},
where a spherical particle (radius $R$) self-propels due to a surface slip velocity~\cite{Magar03},
$u(\alpha) = \tfrac{3}{2}u_0\sin \alpha \left(1+\beta \cos\alpha\right)$, where $\alpha$ is the polar angle with respect to the particle's axis (see Supplementary Fig.~S1), $u_0$ is the free swimming speed of the particle and
$\beta$ defines the hydrodynamic nature of the swimmer: when $\beta<0$ the swimmer is a pusher, while $\beta>0$ corresponds to a puller.
To stop the particles penetrating each other and the wall, we employ a short range repulsive potential~\cite{supplement}. The aligning field is modeled as an external torque $T=B\sin(\psi)$, where  $\psi$ is the angle between the wall normal and the swimmer direction (Fig.~\ref{model}A). Using a fluid viscosity $\mu$ and scaling by a hydrodynamic torque, we define a dimensionless strength parameter $B^*=B/6\pi\mu u_0R^2$where $u_0=10^{-2}$, $R=8$ and $\mu = 1/6$ were chosen. The dynamics is characterised by a Reynolds number $\mathrm{Re}=u_0R/\mu$, for numerical convenience we used $\mathrm{Re}\approx 0.5$. The formation and stability of the assemblies was checked for $\mathrm{Re} \approx 0.05$~\cite{supplement} rendering the results valid for $\mathrm{Re}\ll 1$ microswimmers. {\color{black} The particles are {\color{black}suspended} between two flat walls (separation $\sim 6R$)~\cite{supplement}.}

\begin{acknowledgments}
We are very grateful to Hamid Kellay for careful reading of the manuscript. Z.S. and J.S.L. acknowledges support from IdEx (Initiative d'Excellence) Bordeaux and computational resources from Avakas cluster. A.W. acknowledges support by the French National Research Agency through Contract No. ANR-13-IS04-0003.
\end{acknowledgments}

\bibliography{ref}

\clearpage
\onecolumngrid

\onecolumngrid
\newpage

\makeatletter 
\def\tagform@#1{\maketag@@@{(S\ignorespaces#1\unskip\@@italiccorr)}}
\makeatother

\makeatletter \renewcommand{\fnum@figure}
{\figurename~S\thefigure}
\makeatother

\setcounter{equation}{0}
\setcounter{figure}{0}

\begin{center}
  {\Large \bf Supplementary information for Hydrodynamic assembly of active colloids: chiral spinners and dynamic crystals}
\end{center}

\medskip


\section{Simulation model}
In this work we modelled the hydrodynamics of the active colloids using lattice Boltzmann (LB) simulations. The motile particle is based on a squirmer model~\cite{lighthill52}. The squirmer model does not explicitly deal with phoretic interactions, but the particle is rendered motile by continuous slip velocities over the particle surface, which arise from the differing materials properties at the surface.  The tangential slip velocity at the particle surface is written as~\cite{Magar03}
\begin{equation}
u(\alpha)=B_1\sin \alpha + B_2 \sin \alpha  \cos \alpha .
\label{squirmer}
\end{equation}
where $\alpha$ is the polar angle with the respect to the particle's axis (see Fig. S1). The parameters $B_1$ and $B_2$  give the velocity of a free particle in the bulk as $u_0= 2 B_1/3$ and the squirming parameter $\beta={B_2}/{B_1}$. The $\beta$ defines the nature of the swimmer: when $\beta<0$ the swimmer is a pusher and when $\beta>0$ the swimmer is a puller.

\begin{figure}[tbhp!]
\centering
\includegraphics[width=0.9\textwidth]{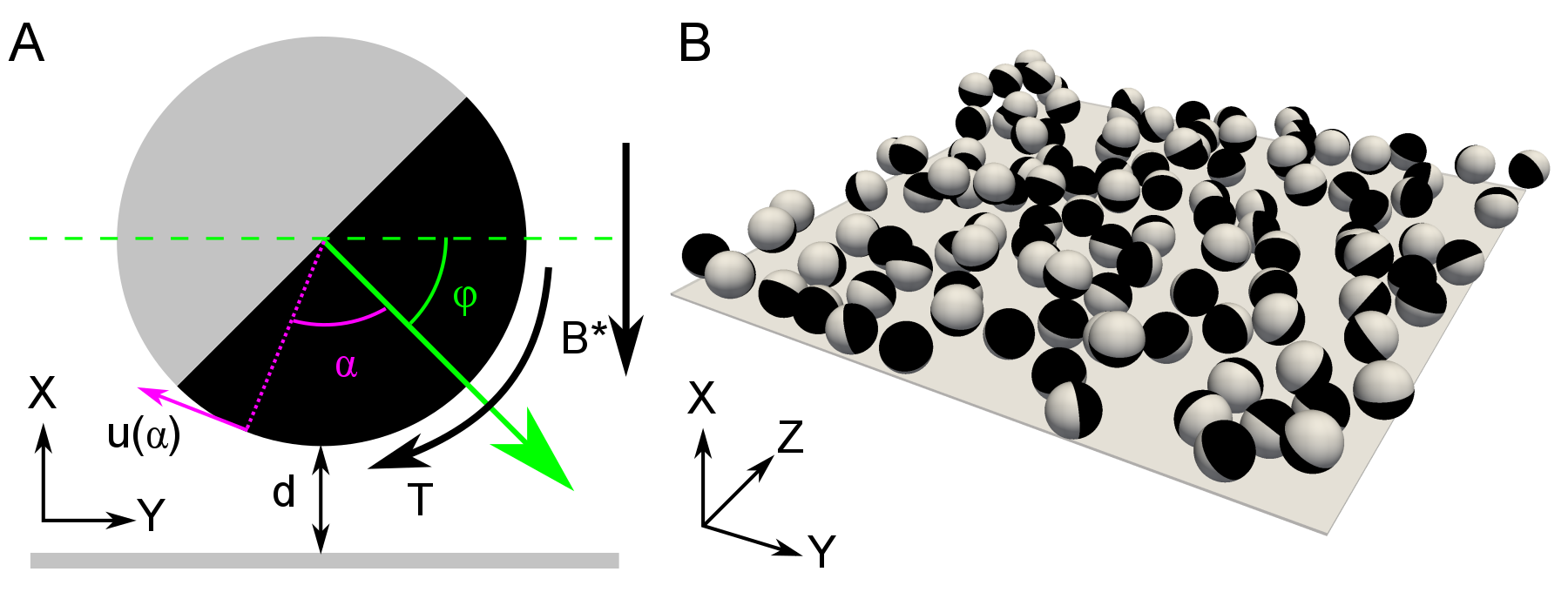}
\caption{\footnotesize \textbf{A schematic showing the simulation set-up.} (\textbf{A}) The green arrow indicates the particle axis as well as the swimmer orientation. $\varphi$ is the angle between the particle direction and the wall plane. An external field $B^*$ is applied to create a torque $T$ which turns the swimmer towards the wall normal, as shown by the black arrow. (\textbf{B}) The particles are initially positioned randomly above the surface.}
\label{schematic}
\end{figure}

In the LBM a no-slip boundary condition at solid-fluid interface can be realised by using a bounce back on links method~\cite{ladd1,ladd2}.
In order to simulate the squirming motion, the boundary condition at the particle surface is modified to include the surface slip flow~\cite{ignacio1,ignacio2}.

To stop the particles penetrating the wall, we employ a short range repulsive potential
\begin{equation}
  V(d) = \epsilon\left(\frac{\sigma}{d}\right)^\nu
  \label{soft}
\end{equation}
which is cut-and-shifted by
\begin{equation}
  V_W(d) = V(d) - V(d_c) - (d-d_c) \frac{\partial V(d)}{\partial d}\mid_{d=d_c}
  \label{repulsion}
\end{equation}
to ensure that the potential and resulting force go smoothly to zero at the interaction range $d_c= 1.2$ in simulations units (the values of below parameters are in simulation units) corresponding to repulsion range of $0.15R$. The $d$ is defined as the distance between the particle bottom and the surface (see Fig. S1). The $\epsilon = 0.6$ and $\sigma = 1.0$ are constant in the reduced units of energy and length, respectively. The $\nu = 12$ controls the steepness of the repulsion.

{\color{black}
The Fig. S1 gives a schematic of our system. {\color{black} The particles are initialised randomly between the two walls}. We apply an external torque to direct the particles towards the bottom wall (Fig.~S1A), 
\begin{equation}
\mathbf{T} = B\mathbf{n} \times \mathbf{a}   ,
  \label{torque}
\end{equation}
where the field direction $\mathbf{n}$ is along the surface normal and $\mathbf{a}$ is the particle axis. Its strength is expressed through the viscous torque, $B=6 \pi \mu u_0 R^2 B^*$, with a dimensionless parameter $B^*$, and through the angle between the unit vectors (see Fig.~1A in the main text), defined by $|\mathbf{n} \times \mathbf{a}|= \sin\psi$.}  In the simulations we used a viscosity $\mu = 1/6$, a spherical particle of radius $R=8$ and fixed the unperturbed bulk swimming speed $u_0=0.01$, {\color{black} which give the Reynolds number $\mathrm{Re}=u_0R/\mu\approx 0.5$ (see section~\ref{re_eff} for comparison with $\mathrm{Re}\approx 0.05$)}. For the phase diagram (Fig.~6A in the main text) we carried out our simulations using $N=90$ particles in a rectangular simulation box with the size of $48\times 240\times 240$, with a no-slip wall at $x=0.5$ and $x=48.5$ and periodic boundary conditions along $Y$ and $Z$. In ordered to simulate large number of independent states, a smaller system of $N=30$ with a box of $48\times144\times144$ was used to study the chiral transfer in Fig. 3 in the main text. In both cases the area fraction corresponds to approximately 30\%.

\begin{figure}[tbhp!]
\centering
\includegraphics[width=1\textwidth]{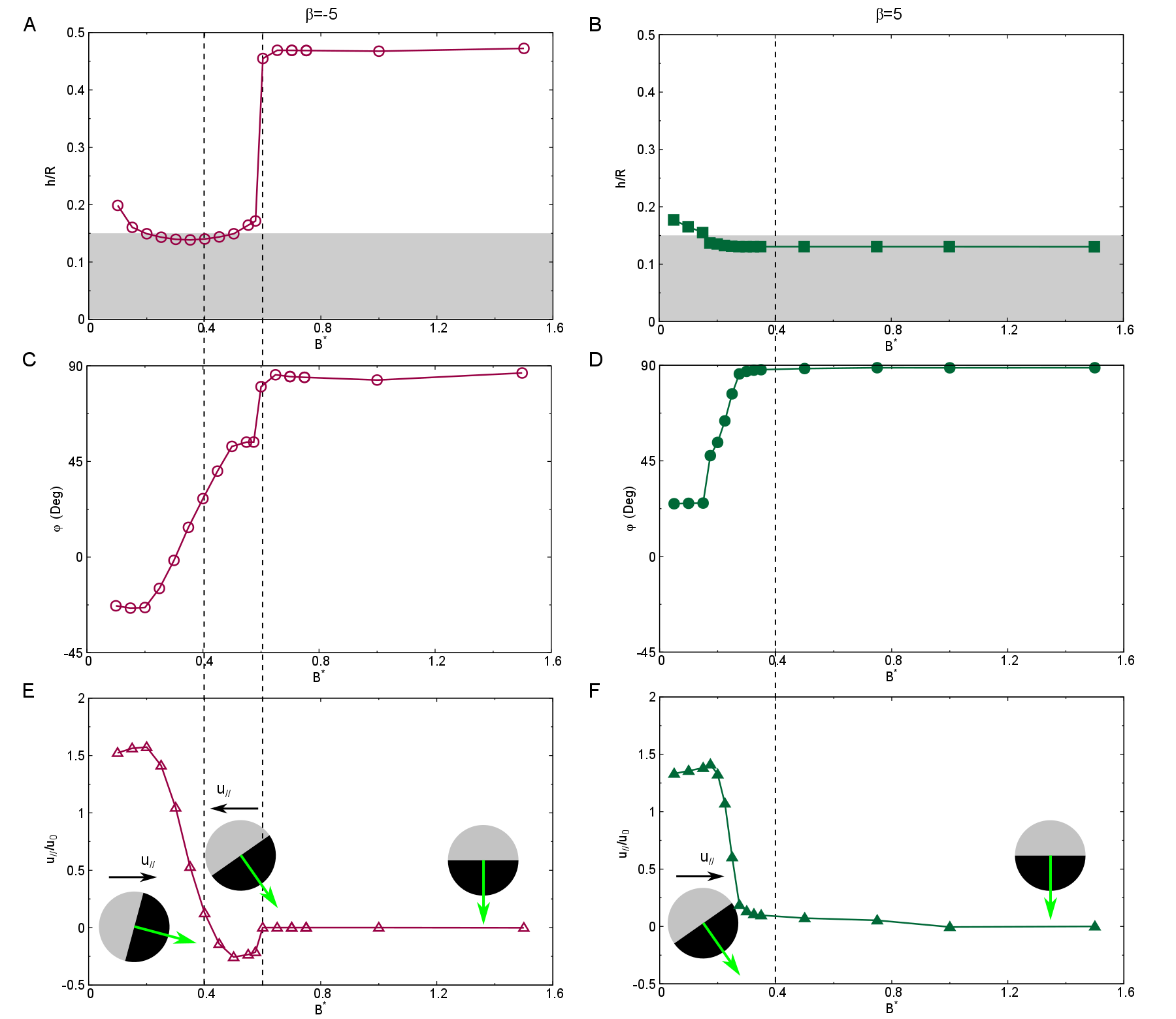}
\caption{\footnotesize \textbf{Dynamics of a single swimmer directed by an external torque near the wall.} (\textbf{A}, \textbf{C} and \textbf{E}) A $\beta=-5$ pusher, the gap size $d$, inclination angle $\varphi$ and the normalised swimming $u_{//}/u_0$ along the wall as a function of the external aligning field $B^*$. (\textbf{B}, \textbf{D} and \textbf{F}) A $\beta=+5$ puller. The shading indicates the range where a soft repulsion is added to avoid the particle penetrating the wall. The dashed lines in the left panel indicate the region where the pusher swimmer reverses its velocity along the wall.}
\label{single}
\end{figure}

\section{Dynamics of a single swimmer near a surface when the orienting field is applied}

Detailed theoretical calculations~\cite{ishimoto13} and lattice Boltzmann simulations~\cite{shen18}, have shown that a single swimmer can be hydrodynamically trapped by the flat wall when $|\beta| \gtrsim 4$, with a well defined gap size $d$ and inclination angle $\varphi$ (Fig.~S\ref{schematic}A). In a steady state the pullers face towards the wall ($\varphi>0$), while pusher point towards the bulk ($\varphi>0$). In Fig.~S\ref{single} the dynamics of a $\beta = -5$ pusher and $\beta = +5$ puller near the wall is presented, when the external torque $B^*$ is applied to orient swimmer direction towards the wall. For very small $B^*$, the wall-particle hydrodynamic interaction dominates. The particles swim near the wall with a steady state $d/R\approx 0.2$ (Fig.~S\ref{single}A, B) in agreement with previous calculations~\cite{ishimoto13} and simulations~\cite{shen18}. The $\varphi$ is about $-25^\circ$ and $25^\circ$ for pusher and puller respectively (Fig.~S\ref{single}C, D), and the translational velocity is increased compared to the bulk, with the value approximately of $1.5u_0$ (Fig.~S\ref{single}E, F), as expected~\cite{lintuvuori16}. When $B^*$ is increased, the external torque compete with the wall-particle hydrodynamic interactions, leading the swimmer direction orient towards the wall. $\varphi$ increases (Fig.~S\ref{single}C, D) and changes sign from negative to positive around $B^*=0.3$ when $\beta=-5$ (Fig.~S\ref{single}C). Further increasing $B^*$, leads to the swimmer orientation perpendicular to the wall, $B^*>0.6$ for a pusher (Fig.~S\ref{single}C) and $B^*>0.4$ for a puller (Fig.~S\ref{single}D), and the particles stop moving (Fig.~S\ref{single}E, F). Generally, the particle velocity decreases when $B^*$ is increased (Fig.~S\ref{single}E, F). 
However, within a certain range of $B^*$ ($0.4<B^*<0.6$) a pusher reverses its swimming direction (between the dashed lines in Fig.~S\ref{single}E). The particle can swim opposite to the swimmer direction and the speed can reach about $0.3u_0$. 

\section{Dynamics of a swimmer pair when the orienting field is applied}

\begin{figure}[tbhp!]
\centering
\includegraphics[width=1\columnwidth]{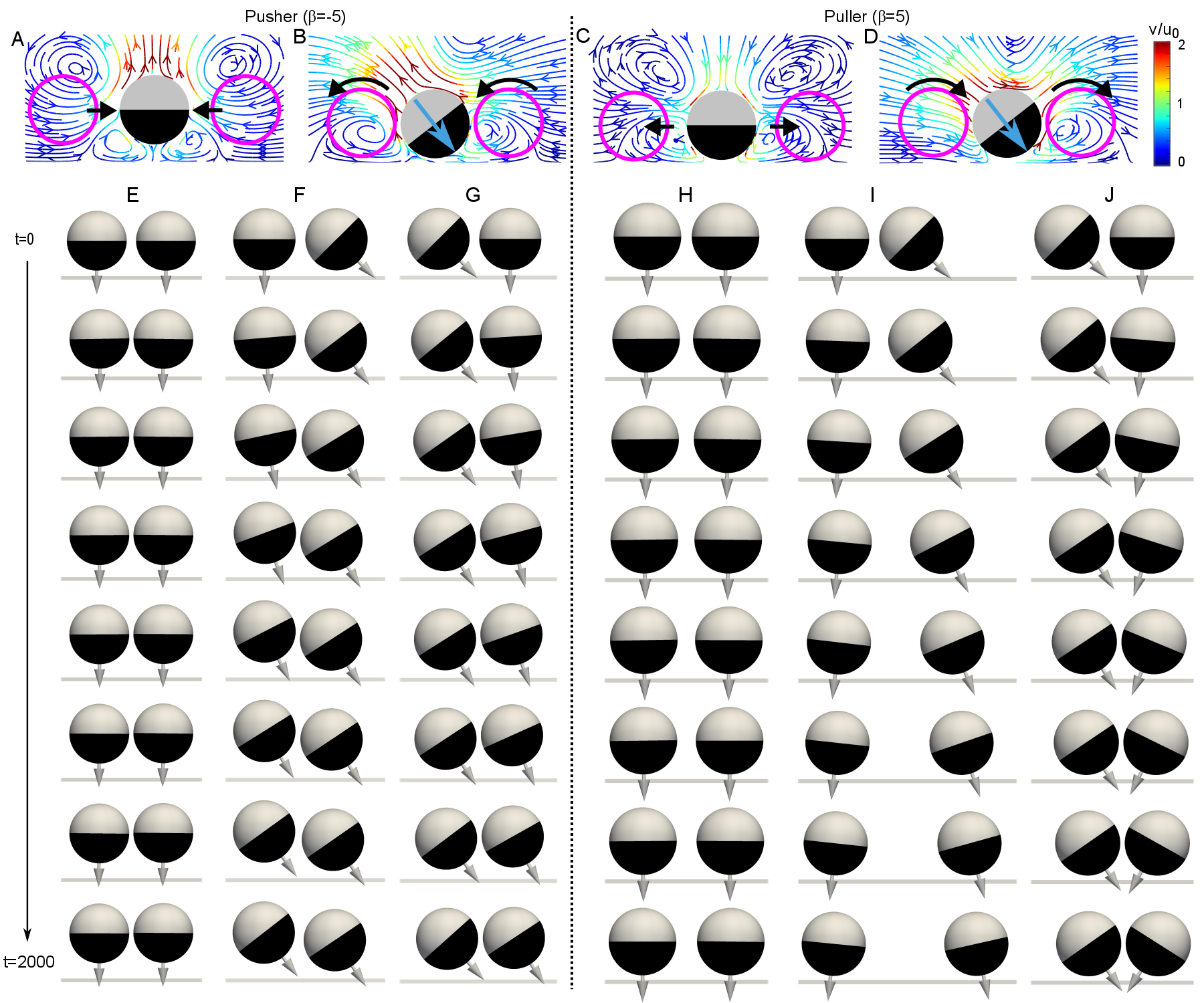}
\caption{ \footnotesize \textbf{Hydrodynamic interactions between two swimmers when the aligning field is turned on.} The flow field in the plane across the particle center and vertical to the wall, is shown in (\textbf{A-D}). (\textbf{A}) For a single pusher ($\beta=-5$) and a strong field ($B^*=1.5$), the imaginary particles (magenta circles) are attracted to the central particle. (\textbf{B}) For a single pusher ($\beta=-5$) and a moderate field ($B^*=0.5$),
the adjacent particles will be oriented by the vortices created by the swimmer, along the direction indicated by the black arrows. (\textbf{C}) For a single puller ($\beta=+5$) and a strong field ($B^*=1.5$), the adjacent imaginary particles (magenta circles) will be repelled. (\textbf{D}) For a single puller ($\beta=+5$) and a moderate torque ($B^*=0.2$). The imaginary particles will be oriented by the flow field, to direction presented by the black arrows. The time evolution (from 0 to 2000 time steps) of two interacting swimmers (\textbf{E-J}):
(\textbf{E}) When a strong field ($B^*=10$) is applied, two pushers are attracted. 
(\textbf{F}) and (\textbf{G}) Upon the application a moderate field ($B^*=1.5$), the two pushers synchronise their orientations.
(\textbf{H}) When strong field ($B^*=10$) is applied two pullers are repelled. 
(\textbf{I}) and (\textbf{J}) at moderate field ($B^*=1.5$), two pullers are oriented opposite directions. They repel each other in downstream case \textbf{I}, while in upstream case \textbf{J} they get bound state. \label{pair}}
\end{figure}

When two active particles are close, the self-generated flows create a particle-particle hydrodynamic interactions which affect the mutual orientations of the swimmers. To analyze how the particles influence each other, we plot the flow field produced by a single particle near a wall for various strengths of the aligning field (Fig.~S\ref{pair} A-D). First, we consider a pusher. When a strong external torque is applied, the particle is perpendicular to the wall. In this case the flow field creates an attraction of the surrounding particles (Fig.~S\ref{pair}A). To test this, we placed two pushers close to each other as shown in top panel in Fig.~S\ref{pair}E. The time series presented in Fig.~S\ref{pair}E, show that the particles retain their orientation perpendicular to the wall, but move closer to each other, thus validating the attraction. When the external torque is reduced and it is comparable to hydrodynamically induced particle-particle interactions, the swimmer direction is out of vertical (Fig.~S\ref{pair}B). 
In this case, there is two counterclockwise rotating vortices at both sides of the particle, which will rotate the surrounding particles also counterclockwise, synchronising with the source particle. In Fig.~S\ref{pair}F and G, we show the corresponding time series where a swimmer initially vertical is reoriented by the source particle. Finally we observe that the two swimmers have same orientation (Fig.~S\ref{pair}F and G).

Next we turn to pullers. Upon the application of a strong field, when the isolated swimmer remains perpendicular to the surface, the flow field of a puller (Fig.~S\ref{pair}C) gives a rise to an opposite effect to a pusher: a vertical puller repels of the surrounding particles, as can be seen from the time series in Fig.~S\ref{pair}H. When the external torque is reduced, the hydrodynamic torques compete with the external torque (Fig.~S\ref{pair}D). Contrary to the pusher, a puller 
produces two clockwise rotating vortices on the both sides of the particle (Fig.~S\ref{pair} D). This leads to a directional interactions, where a particle placed downstream is rotated away from the source swimmer and subsequently repelled, while up-stream particle is rotated towards the source swimmer, leading to the formation of hydrodynamically bound particle dimer, as shown in the time series of Fig.~S\ref{pair}I and J.

\if{
\section{Dynamics of a swimmer pair when the orienting field is applied}
To realise the a generic route for the dynamic self-assembly, the particles are oriented towards the surface by an external aligning field.
When the external torque, which could be realised for example by gravity for top heavy particles or by a constant magnetic field for weakly magnetic active particles~\cite{palacci2013living}, is comparable to the wall-particle hydrodynamic interaction, an isolated particle can swim along the surface with a well defined inclination angle $\varphi$ and distance $d$ (Fig.~S\ref{single}).
Pusher reverses it swimming direction (Fig.~S\ref{single}A) and when moving along $y$ in the Fig.~\ref{strategy}a in the main text, it creates two anticlockwise rotating vortices at both sides of the particle (Fig.~S\ref{pair}). This flow rotates the adjacent particles, leading to the synchronisation of the orientations (Fig.~\ref{strategy}A). The long-range hydrodynamic interactions are attractive~\cite{singh2016universal} (see Fig.~\ref{strategy} in the main text and Fig.~S\ref{pair}), leading to an accumulation of the particles (Fig.~\ref{strategy}a in the main text).
Opposite to pushers, a puller creates a clockwise rotating flow at both sides of the particle (fig.~\ref{strategy}B and supplementary figure~S\ref{pair}).
This leads directional interactions, where a particle placed up-stream is attracted but in down stream repelled, giving a rise to hydrodynamically bound small clusters (Fig.~\ref{strategy}b).
}\fi

\section{Distribution of the chiral spinners}
Here we discuss the distribution of the active colloids engaged in small clusters and, in particular, the imbalance of the particles in the counterclockwise and clockwise spinners, $I = N^--N^+$ . According to Fig. 3B in the main text, the probability of monomers in a $N=30$ sample is $q_0\approx8\%$, that of particles engaged in a left-handed or right-handed dimer $q_{d\pm}\approx18\%$, and for trimers $q_{t\pm}\approx28\%$.

In the absence of interactions between neighboring clusters, the probability distribution of the swimmers then is given by the multinomial distribution,
\begin{equation}
 Q = \frac{1}{\mathcal{N}}  \frac{N!}{n_0!n_{d+}!n_{d-}!n_{t+}!n_{t-}!}
      q_0^{n_0} q_{d+}^{n_{d+}}  q_{d-}^{n_{d-}}  q_{t+}^{n_{t+}}  q_{t-}^{n_{t-}}  ,
\end{equation}
where $n_{d+}$ and $n_{d-}$ are even numbers, and  $n_{t+}$ and $n_{t-}$ are multiples of 3.  The sum of the $n_i$ gives $N$, and $\mathcal{N}$ is a normalization constant.

For large numbers the above expressions can be simplified by using Stirling's formula $\ln n!=n\ln n-n$ and reducing the distribution function to Gaussian.  As in the main paper we use the total number of right-handed and left-handed squirmers, $N^\pm=n_{d\pm}+n_{t\pm}$. For the imbalance $I = N^--N^+$ one finds
\begin{equation}
   P(I) = \frac{e^{-I^2/2\Delta I^2}}{\sqrt{2\pi}\Delta I},
   \label{S2}
\end{equation}
with the width
\begin{equation}
\Delta I^2 = \left[(2q_{d+})^2 + (2q_{d-})^2 + (3n_{t+})^2 + (3n_{t-})^2 \right] N .
\end{equation}
Noting that clockwise and counterclockwise are equally probable, one finds
\begin{equation}
\Delta I =  \sqrt{2(4q_d^2+9q_t^2)N} ,
\end{equation}
With the measured cluster probabilities of Fig. 3B for $N=30$, the numerical factor reads $\Delta I \approx 6$. When fitting the actual distribution function resulting from $Q$ by a Gaussian, we obtain a slightly smaller value, $\Delta I \approx 5$.

Thus one would expect that the imbalance obeys a narrow Gaussian distribution $P(I)$, whereas the simulation results shown in Fig. 3D, reveal that $P(I)$ is roughly constant.

\begin{figure}[tbhp!]
\centering
\includegraphics[width=1.0\textwidth]{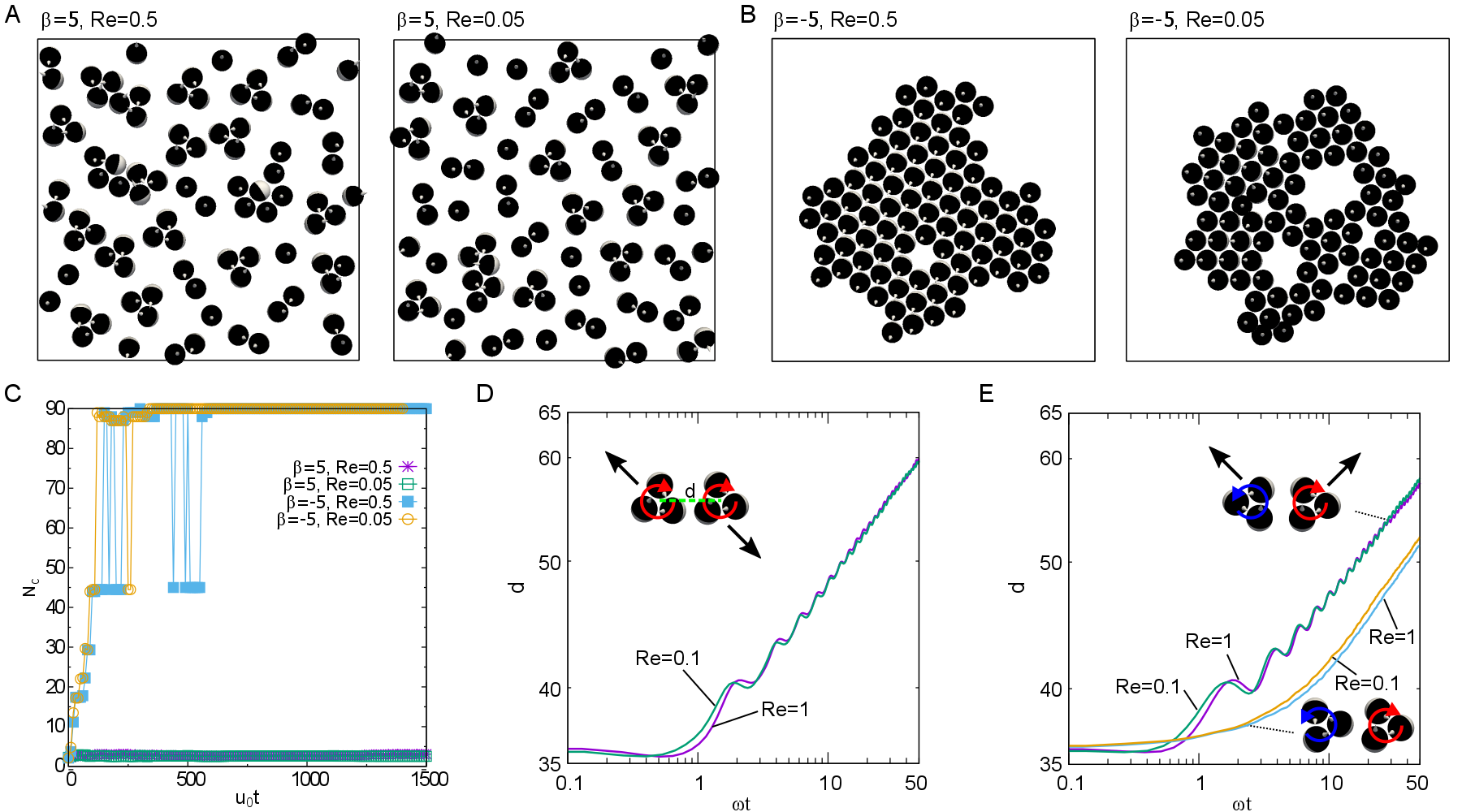}
\caption{\footnotesize \textbf{Stability of the self-assembled structures when Reynolds number (Re) is varied.} (\textbf{A}) Observed spinner structures from the simulations with strong pullers ($\beta=+5$) when $\mathrm{Re}\approx 0.5$ as in the main text (left panel) and $\mathrm{Re}\approx 0.05$ right panel, with $B^*=2$. (\textbf{B}) Observed motile aggregate structures from the simulations with strong pushers ($\beta=-5$) when $\mathrm{Re}\approx 0.5$ as in the main text (left panel) and $\mathrm{Re}\approx 0.05$ right panel, with $B^*=6$. (\textbf{C}) The cluster size $N_c$ as a function of time. The simulations were carried out with $N=90$ particles. The time evolution of the distance $d(t)$ between (\textbf{D}) co-rotating and (\textbf{E}) oppositely rotating spinner pairs, when the Reynolds number is varied (see text for details). \label{re_effect}}
\end{figure}

{\color{black}
\section{The effects of the Reynolds number\label{re_eff}}
The microswimmers move slowly in a viscous media. Thus the ratio between inertial and viscous forces is small, characterised by a low Reynolds number $\mathrm{Re} << 1$. In our simulations, the wall induced hydrodynamic interactions slow down the particles and mixing characterised by the magnitude of the squirmer parameter $|\beta|$ are expected to dominate. To achieve reasonably numerical convergence of our simulations, we chose a reasonably high bulk swimming speed $u_0 = 1.0\times10^{-2}$ in simulations units, this with the particle radius $R=8$ and fluid viscosity $\mu = \tfrac{1}{6}$, gives a Reynolds number for a single particle in the bulk $\mathrm{Re}=\tfrac{u_0R}{\mu}\approx 0.5$. To check the validity of our predictions for $\mathrm{Re}<<1$ we carried out further simulations where the swimming speed was reduced $u_0=10^{-3}$, giving $\mathrm{Re}\approx 0.05$.  These are compared to the structures observed with $\mathrm{Re}\approx 0.5$ in Fig.~S\ref{re_effect}. Both in the case of strong puller spinners (Fig~S\ref{re_effect}A) and  dynamic crystal with strong pusher $\beta = - 5$ (Fig.~S\ref{re_effect}B), we see similar structures between the two Reynolds numbers. This is further supported by looking the time development of the average cluster size $N_c$, which shows very similar dynamics between the two Reynolds numbers considered as shown in Fig.~S\ref{re_effect}C.}

The spinning frequency $\omega$ of a hydrodynamically bound spinners scales linearly with $u_0$ and thus with the particle Reynolds number. For the simulations above we find $\omega \approx 2\pi\times 10^{-4}$ and $\omega\approx 2\pi\times 10^{-5}$ for $\mathrm{Re}\approx 0.5$ and $\mathrm{Re}\approx 0.05$, respectively. Using the particle diameter as the spinner size, we find corresponding Reynolds numbers of the spinners as $\mathrm{Re} = \omega(2R)^2/\mu\approx 1.0$ and $\mathrm{Re}\approx 0.1$. The hydrodynamic repulsion between the spinner arises from the hydrodynamic pumping from the indiviudal particles and remains unaffected by the choice of the Reynolds number (Fig.~\ref{re_effect}D and E).

\begin{figure}
\centering
\includegraphics[width=0.8\columnwidth]{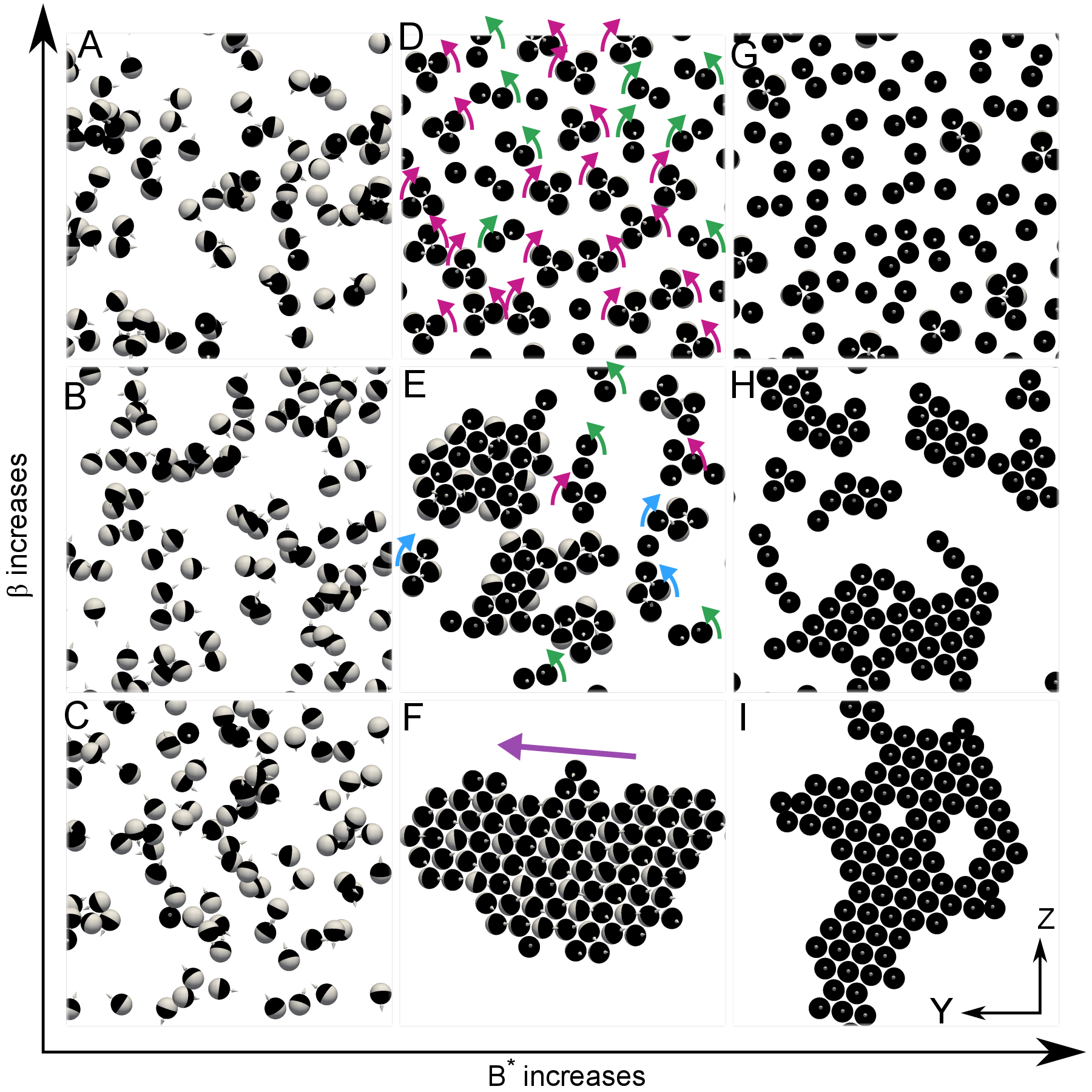}
\caption{ \footnotesize {\bf Typical configurations observed for different values of $\beta$ and $B^*$.} The parameters for each snapshot: (\textbf{A}) $\beta=+5$, $B^*=0$. (\textbf{B}) $\beta=0$, $B^*=0$. (\textbf{C}) $\beta=-5$, $B^*=0$. (\textbf{D}) $\beta=+5$, $B^*=2$. (\textbf{E}) $\beta=+2$, $B^*=1$. (\textbf{F}) $\beta=-5$, $B^*=4$. (\textbf{G}) $\beta=+5$, $B^*=3$. (\textbf{H}) $\beta=+2$, $B^*=3$. (\textbf{I}) $\beta=-5$, $B^*=8$. All simulations were carried out with 90 particles, with an area fraction equal to $31\%$. (\textbf{A-C}) Gas-like state, particles move randomly. (\textbf{D-F}) Dynamic cluster state, the arrows indicate the moving directions of spinners and aggregates. (\textbf{G-I}) Stationary cluster state, cluster size increases with decreasing $\beta$. \label{phase_config}}
\end{figure}

\begin{figure}[tbhp!]
\centering
\includegraphics[width=0.8\columnwidth]{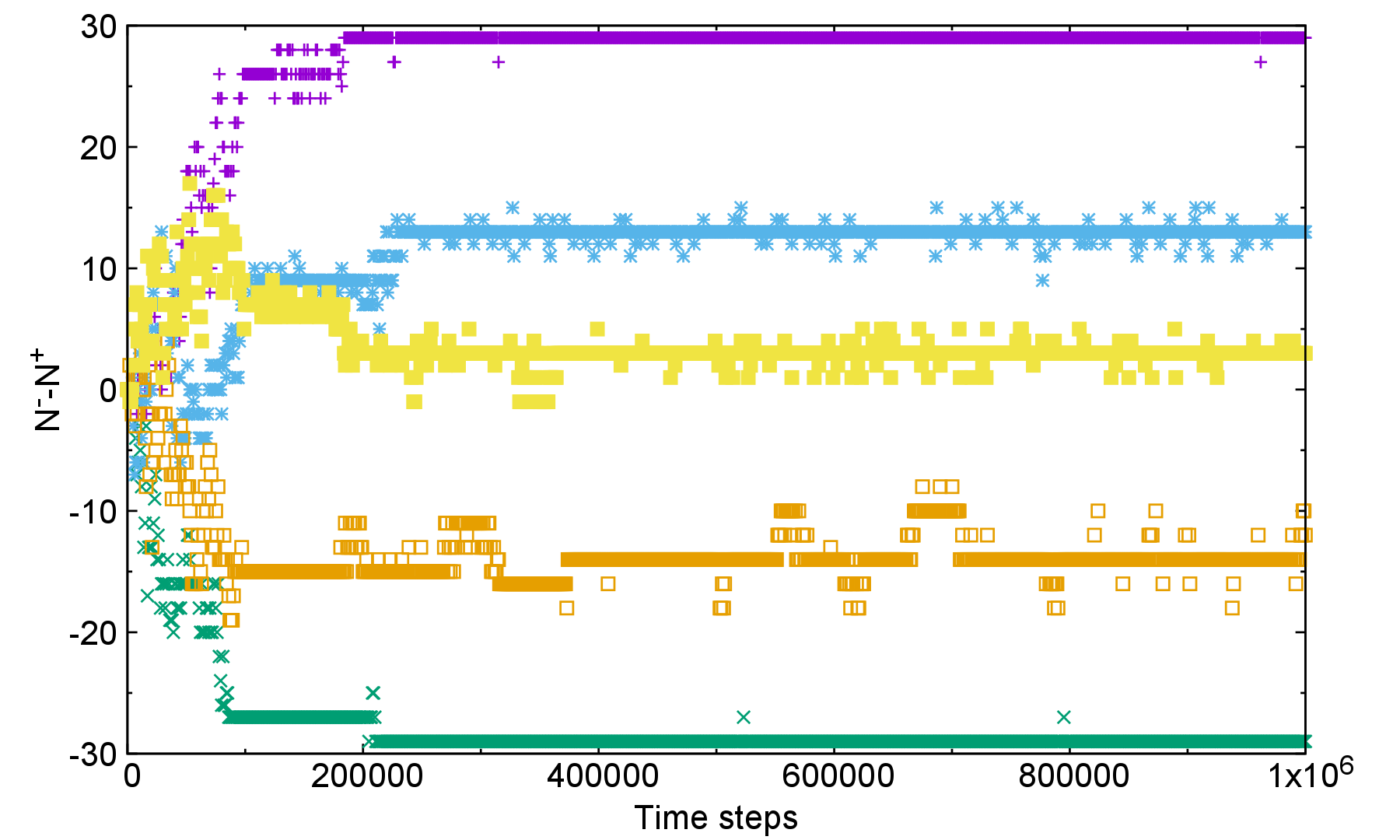}
\caption{\footnotesize \textbf{The time evolution of the imbalance.} The imbalance $I=N^--N^+$ between particles in counterclockwise and clockwise turning spinners as a function of time, observed from 5 independent simulations with $N=30$ particles, $\beta = +5$ and $B^*=2.0$ corresponding to the Fig. 3 in the main text. \label{imbalance_evo}}
\end{figure}



\clearpage

\section*{Movies}

Movie 1: Dimers and trimer spinners formed in a suspension of $\beta = +5$ pullers with $B^*=2.0$.\\
Movie 2: Homochiral and racemic mixture self-assembly ($\beta = +5$ and $B^* = 2.0$). After 10s the movie playback speed is increased 40 times. \\
Movie 3: Weak $\beta = + 1$ puller, neutral $\beta = 0$ and weak $\beta = -1$ pusher, with a weak aligning field $B^* = 0.75$ ($N=300$ particles and $48 \times 640 \times 640$  computational domain corresponding to an area fraction of $\phi\approx 15$\%).\\
Movie 4: A random moving aggregate formed by pushers. $\beta=-5$ and $B^*=2.0$.\\
Movie 5: A rotating aggregate formed by pushers. $\beta=-5$ and $B^*=4.0$. \\
Movie 6: A translating aggregate formed by pushers. $\beta=-5$ and $B^*=4.0$.\\
Movie 7: A translating aggregate formed by pushers with completely synchronized internal orientations.\\
Movie 8: 5 tests of the transfer of chirality from a trimer to initially achiral dimer. $\beta = + 5$ and $B^* = 2.0$
\\
\\
Movies 1 and 4-7 were carried out with 90 particles and movie 2 with 72 particles, in the computational domain of $48 \times 240 \times 240$. All the movies use bottom view and in the movies 1 and 4-7 $2 \times 2$ computational domains are shown.

\end{document}